\documentstyle[12pt]{article}
\begin{document}
\title{About model of the Universe with accelerated movement of the time}
\author{ Wladimir B. Belayev\\
\normalsize Center for Relativity and Astrophysics,\\
\normalsize Box 137, 194355, Sanct-Petersburg, Russia\\  
\normalsize e-mail: vladter@ctinet.ru}

\maketitle  

\begin{abstract}
 Cosmological model based on metric of Fridmann-Robertson-Walker
 with permanent size and acceleration of  time is considered.
 The problem of the dark matter is analyzed within this model .
\end{abstract}

  It is supposed at present that the Universe has arisen about $1.5\times 10^{10}$ years 
ago by means of the explosion of concentric mass of the matter from the dot, and a 
redshift  of the spectra of the distant galaxies is caused by their motion after this 
explosion. At the same time, no one can  tell about it as a provable fact. There are 
not observations or experimental results at present providing that redshift is caused by 
Universe's extension. 

 Milne [1,2] proposed other approaches to solution of cosmological 
problems. He supposed in theory with a cosmic scale of the time, that the redshift is 
caused by acceleration of  time, and, it means, that the time was delayed in the 
past. Elements of this model of Universe are in [3,4]. Such approach is 
considered in present paper. Microvawe cosmic background and problem of the dark 
matter are investigated within this model.

 If dinamics of gas is used for discription of the Universe, the solution of Einstein's 
equation with cosmological constant $\Lambda$ Ref. [2,5]
\begin{equation}
G_{ik}-0.5g_{ik}G=-\kappa T_{ik}-\Lambda g_{lm}                                                                                           
\end{equation}
in spherical space is metric of Fridmann-Robertson-Walker [4] eq. (2) .
\begin{equation}
ds^2 =Q^2\left( t\right) dt^2-R^2\left( t\right) /c^2( \left[ dr^2+r^2d\theta ^2+r^2\sin^2\theta d\phi ^2\right] /\left[ 1+kr^2/4\right] ^2)   
\end{equation}
With coefficients corresponding  to this metric , equations (1) reduce [5] to eq. (3), 
\[
\kappa c^2\rho = 3/R^2\left( kc^2+ \left( \dot R\right) ^2 /Q^2 \right) -\Lambda   ,
\]                                      
\begin{equation}
 \kappa c^2p =-2\ddot R/\left( RQ^2\right) -\left( \dot R\right) ^2/\left( RQ\right) ^2+ 2\dot R \dot Q/\left( RQ^3\right) -kc^2/R^2+\Lambda   
\end{equation}
where $\rho$ is density, $p$ is pressure, in homogeneous space.

A redshift is 
\[
\delta =d\lambda /\lambda ,
\]
where $\lambda$ is the length of the wave in the point of the radiation , $\lambda +d\lambda$ is length of wave in a point of  observation. Equation of the geodesy line [5] for metric (2) have solution 
\[
dr/dt =-c\left( 1+kr^2 /4\right) Q/R  .
\]

Then following [5] we obtain eq. (4),
\begin{equation}
\delta =R_oQ_i /\left( R_iQ_o\right) -1
\end{equation}
where $R_o$ , $Q_o$ are quantities of coefficients $R$, $Q$ in moment of time $t_o$ of signal's 
receipt in point $O$; $R_i$ , $Q_i$  quantities of these coefficients in moment of time $t_i$  of light radiation in point $O_i$. Both these points have fixed coordinates. 
 
 Now the model of the Universe will be considered.  Let's assume, that the size of the Universe is permanent : $R=const$ , and distance $R_N$ to the most distant objects being observed at present time is such that $R_N\ll R$.  In this case equations (3) are
\[
\kappa c^2\rho = 3kc^2/R^2-\Lambda ,
\]                         
\[
 \kappa p =-kc^2/R^2+\Lambda .     
\]

Basing on the model of a Big Bang Gamov predicted Microvawe Cosmic 
Background (MCB). Temperature of MCB was predicted fairly exactly by principle of 
a hot model of the first period of the Universe's evolution [6,7]. However, a process of 
nuclear synthesis corresponding to the model of the expanding Universe can be also if the size of the Universe is permanent. It is possible if the cooling of the 
light radiation takes place owing to acceleration of the time. This is insured  from that the 
type of nuclear reaction is determined by the photon energy. In this case correlation 
between the density of matter and radiation after recombination can be the same 
as in case of the expanding Universe, but the number of photons of the MCB in the unit of 
the volume is being constant after recombination. Thus the first moment of Universe's 
existence may be not a Big Bang , but a simultaneous origin of reducing 
blend of matter and radiation with high temperature in a whole space of the Universe.
 
Now trend of the time in this space-time will be considered. As ensured from (2), 
correlation between a small interval of time $dt_i$ between events having taken place in  
point $O_i$ beginning since moment of time $t_i$  and interval of time $dt_o$ , in  which they 
observed in point $O$ beginning since moment of time $t_o$ , is   
 \[  
 Q_idt_i=Q_odt_o  .
\]  
Having set 
\[
t^*=\left( t_o-t\right) Q_o  ,
\]
\begin{equation}   
Q^*\left( t^* \right) =Q\left( t\right) /Q_o 
\end{equation}
we obtain eq. (6) .
\[
t_o^*=0  ,
\]
\begin{equation}   
 Q^*\left( 0\right) =1 
\end{equation}

Now we define time $\tau_{io}$ in which a ray of light must run from point $O_i$ to point $O$ in system of coordinates $\left( O, t_o^*\right)$ eq. (7), 
\begin{equation}   
\tau_{io}=\int_{0}^{t_{io}^*}Q^*\left( t^*\right) dt^*                                                                                 
\end{equation}
where  $t_{io}^*$ is the same interval of time in scale of the time that is coherent with metrics 
(2) . At that 
\[
r_{io}=c\tau_{io} , 
\]
where $r_{io}$  is the distance between points $O$ and $O_i$ .

Now a change of temperature of MCB $T$ will be considered. It is inversely proportional to 
length of wave $\lambda_r$ of the fixed element of the MCB 
\[
T\sim 1/\lambda_r .
\]
We can write down eq. (8),  
\begin{equation}
T= B\lambda_{or}/\lambda_r 
\end{equation}
where  $\lambda_{or}$ is length of wave of the element of the MCB at present , $B$ is a coefficient of 
the proportionality 
\[
B\approx 3K .
\]

As ensured from (4) a relative change of the length wave is (9) . 
\begin{equation}
\delta =Q_i /Q_o-1
\end{equation}
We define now the temperature of MCB in moment, when seen with redshift $\delta$ object in point $O_i$ 
radiated the light. Coefficient $Q_i^*$ corresponding to the 
moment of the radiation is obtained by substituting (5), (6) in (9) and is in eq. (10). 
\begin{equation}
Q_i^*=\delta +1 
\end{equation}
The length of wave of the element of the MCB being considered in moment of time $t_i^*$   is 
\[
\lambda_{ir}=\lambda_{or}/ Q_i^*=\lambda_{or}/\left( \delta +1\right) .
\]
Then eq. (8)  can be rewritten as
\[    
T=B\left( \delta +1\right) .
\]
Thus, for example, temperature of MCB $T\approx 18K$ corresponds to meaning $\delta =5$. It means 
also, that the recombination took place when 
\[
\delta \approx 4000K/3K\approx 1300 .  
\]
  
We supposed that since some moment in the early stage  of Universe's evolution eq. (11),
\begin{equation}
\delta =Ht=Hr/c
\end{equation}
where $H$ is Hubble's constant, is true. And then integral (7) is transformed with (10) to
\[
\delta/H=\int_0^{t^*}\left[ \delta \left( t^*\right) +1\right] dt^* .
\]
We differentiate this equation :
\[
d\delta /H=\left[ \delta \left( t^*\right) +1\right] dt^*,  
\]
and hence can write down
\[
d\delta /\left[ H\left( \delta +1\right) \right] =dt^* .
\]
Having integrated this equation with consideration of initial conditions 
\[
t^*=0, \delta =0
\]
we obtain 
\[
t^*=ln\left( 1+\delta \right) /H  .
\]
Also , it is ensued from this equation and (2) , (5) , (10) , that coefficient $Q\left( t\right)$ of metric 
(2) is
\[
Q\left( t\right)  =exp\left[ \left( t_o-t\right) H\right] ,
\]
which correspond to theory of Milne with cosmic scale of time . 
 
 In considered above examples having been believed $H\approx 75\times 10^{12}year^{-1}$  we obtain that quantity of  redshift $\delta =5$ corresponds to time $t^*\approx 2.4\times 10^{10}year$, and recombination took place about $10^{11}$  years ago.

We investigate now the problem of the dark matter. According to [7], one of 
hypotheses accounting for its presence is existence of large amount of  brown dwarfs, 
which are burnt down white dwarfs. Discrepancy between age of the Universe and the 
time necessary for the formation of the brown dwarfs was being considered as 
shortcoming of this hypothesis,  since the time of the life of the white dwarfs is $10^{10}$  
years and more, according to estimation. Still if the time having passed from the 
beginning of the stars formation exceed by order $10^{11}$  years, this contradiction takes off. 
Indeed, if we assume the mean time to the form multitude of visible stars as 
$\left( 6-10\right) \times 10^9$  years and the dark matter approximately as ten of their masses, 
the time to form the stars is $\left( 6-10\right) \times 10^{10}$  years with constant rate of stars' formation. That corresponds to the time to having passed after recombination obtained above. 
 
At the same time, there is not evidently that an acceleration of time has been always 
linear in accordance with (11) .

The considered model explains homogeneity of MCB angular power spectrum , since 
it admits that the time of  Universe's existence until recombination was being more than 
after its, and photons, in visible field, had interaction until recombination.

As known, the Universe can't be stationary [6], since in this case neighboring galaxies 
should be attract each other. Still mutual attraction may be compensated by the  moment 
in their relative movement. 

Milne's theory with a cosmic scale of the time  is the closest to this model. Yet, 
the model of Milne supposes infinity and homogeneity of the Universe in time and in space. 

This paper is not detailed description of the Universe with acceleration of time and 
permanent size, but it is show that this possibility can't be rejected at present.


\begin{thebibliography}{99}

\bibitem{Milne} E. A.  Milne, {\it Relativity, Gravitation and World-Structure}, Univ. Press, Oxford,
1935; {\it Cinematic Relativity},  Univ. Press , Oxford, 1948.

\bibitem{encycl} A. L. Zelmanov, {\it Cosmology}, Physical encyclopedia , {\bf V2}, Moscow, 1962.                                                              
 
\bibitem{Anton}  I. Antoniadis , C. Bachas , J. Pllis and D.V. Nanopoulos , Phus. Lett. {\bf B11}, (1988), 393; Nucl. Phus. {\bf B328}, (1989), 117; Phus. Lett. {\bf B257} (1991), 278.

\bibitem{Flav} Flavio G. Alvarenca,  Nivaldo A. Lemos ,{\it Dynamical vacuum  in quantum cosmology}, gr-qc/9802029. 

\bibitem{Vittie}, G.C. McVittie, {\it General relativity and cosmology}, Chapman and Hall Ltd. London , 1956. 

\bibitem{Novic}, I.D.Novicov, {\it Evolution of the Universe}, Science, Moscow, 1990.
        
\bibitem{Gurev}, L.E. Gurevich, A. D. Chernin, {\it Origin of galaxies and stars}, Science, Moscow, 1987. 

\end{thebibliography}
\end{document}